\def\beq{\begin{equation}}
\def\eeq{\end{equation}}
\def\blg{\begin{align}}
\def\elg{\end{align}}
\def\beg{\begin{gather}}
\def\eeg{\end{gather}}
\def\bea{\begin{eqnarray}}
\def\eea{\end{eqnarray}}
\def\bed{\begin{displaymath}}
\def\eed{\end{displaymath}}
\def\bef{\begin{figure} \begin{center}}
\def\eef{\end{center} \end{figure}}

\def\nn{\nonumber}

\def\1{\'{\i}}
\documentclass[preprint,pre,aps]{revtex4-1}
\usepackage[utf8]{inputenc}
\usepackage{graphics,graphicx,amssymb,amsmath,dcolumn,bm,url}

\begin{document}
\title{Nonlocal coupling among oscillators mediated by a slowly diffusing substance}
\author{R. L. Viana \footnote{Corresponding author. e-mail: viana@fisica.ufpr.br} and R. P. Aristides}
\affiliation{Department of Physics, Federal University of Paran\'a, 81531-990, Curitiba, Paran\'a, Brazil}

\date{\today}

\begin{abstract}
A general theory is presented for the coupling among nonlinear oscillators mediated by a diffusing chemical substance. We extend a model originally developed by Kuramoto, who supposed that the diffusion characteristic time is much shorter than the oscillator main period, such that diffusion occurs very fast. We eliminate this constraint and consider diffusion to have an arbitrary characteristic time, by solving exactly the diffusion equation using suitable Green functions. We present results in one, two and three dimension, with and without boundary conditions. 
\end{abstract}

\maketitle

\section{Introduction}

In many problems of physical and biological interest we consider nonlinear oscillators (or ``cells'') whose interaction is mediated through a chemical substance which is secreted by the cells and diffuses along the inter-cellular medium, being absorbed by the cells \cite{winfree}. Moreover, the rate of secretion can depend on the cell dynamics, as well as the rate of absorption. In this way the dynamics of the cells are effectively coupled by the diffusing chemical substance, leading to a non-local coupling type which depends on the details of the diffusion process. 

This kind of chemical coupling has been mathematically described by a reaction-diffusion model by Kuramoto leading to non-local couplings \cite{kuramoto95}. In this model the state variables of each oscillator influence the secretion of a chemical substance obeying a diffusion equation. The rate of absorption depends on the local concentration of this substance at each cell position. The spatio-temporal dynamics of such non-locally coupled systems is extremely rich, giving rise to phenomena like power-law spatial correlations \cite{kuramoto96}, chimeras \cite{kuramoto97}, Turing-Hopf mixed mode solutions \cite{battogtokh99}, noisy on-off intermittency with multi-scaling properties \cite{nakao99}, noise-induced transition to turbulence \cite{kawamura07}, and phase turbulence \cite{battogtokh02}. 

Besides the interesting dynamical features of such model, it has been used to describe neuronal networks of non-locally coupled Hodgkin-Huxley equations with excitatory and inhibitory coupling \cite{sakaguchi06}, the transition to bursting synchronization in neuronal networks \cite{viana12}, the interaction among biological clock cells of the suprachiasmatic nucleus \cite{fabio16}. In the latter example it is thought that the synchronization of clock cells is due to their interaction mediated by $\gamma$-aminobutyric acid (GABA) or other neurotransmitters \cite{liu97,liu00}. The transition to synchronization of networks of generic phase oscillators with coupling mediated by a rapidly diffusing substance has been investigated in detail by numerical \cite{batista17} and analytical methods \cite{viana17}. 

In the original version of the Kuramoto model the diffusion characteristic time is much shorter than the oscillator main period, such that diffusion occurs very fast, i.e. the system achieves instantaneously a stationary concentration of the diffusing chemical \cite{kuramoto95}. However, this model cannot be properly applied if the diffusion time is comparable or is much bigger than the oscillator period. In this case one has to consider the diffusion equation and the corresponding Green function. In this paper we solved the diffusion equation for cases in one, two and three dimension, with and without boundary conditions. The resulting interaction kernels can be of interest for modeling of physico-chemical and biological problems in which the coupling among oscillators is mediated by a slowly diffusing substance. 

Let us consider that the dynamical behavior of each ``cell'' is described by a limit-cycle oscillator with average period $T$, we can model with the results of this paper problems in which the interaction is mediated by a substance with typical diffusion time $\sim T$, for example. This seems to be the case in many problems of chemotaxis, which is the ability of cells to move along the direction of concentration gradients of signaling chemicals \cite{veltman08}. 

One of the most investigated organisms with respect to the chemotactic response is the amoeabae {\it Dictyostelium}. In the absence of food the {\it Dictyostelium} cells release signal molecules into their environment, such that they can find other cells and move to create clusters \cite{cai12}. During this process, about $100,000$ {\it Dictyostelium} cells release the chemoattractant cyclic adenosine monophosphate (cAMP) every $6$ min during periods reaching $5$ to $6$ hours after starvation \cite{cai12}. This suggests that the diffusion is very slow compared with internal rhythms of the individual cells.

This paper is organized as follows: in Section II we present the coupling model based on the solution of a reaction-diffusion equation governing the interaction among cell oscillators. In Section III we show the Green function in free space in one, two and three dimensions. The case of finite domains (Dirichlet boundary condition) is considered in Section IV. Our Conclusions are left to the final Section.

\section{Coupling model}

\begin{figure}
\begin{center}
\includegraphics[width=0.8\textwidth,clip]{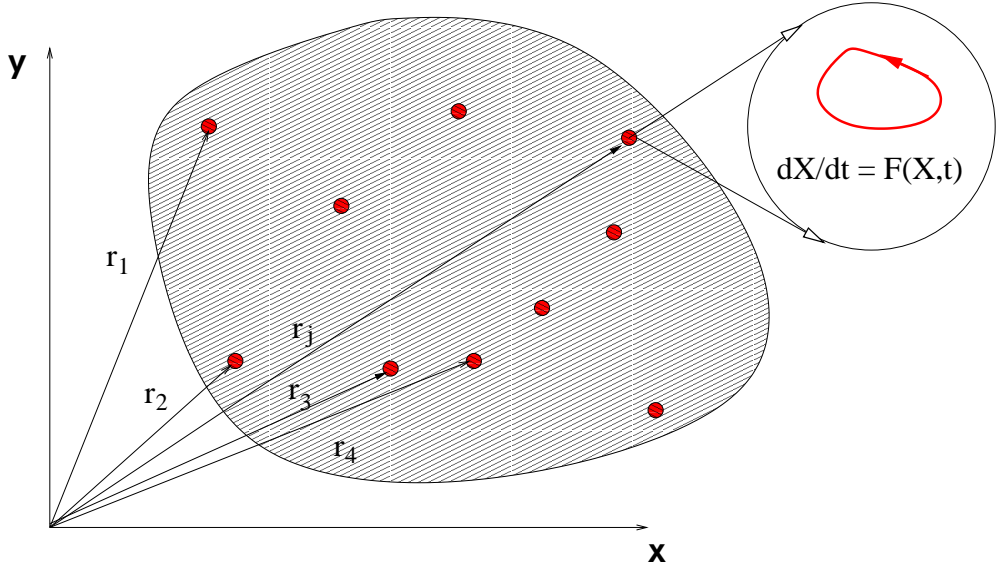}
\caption{\label{chemfig} Schematic figure of the coupled oscillators.}
\end{center}
\end{figure}

In the following we will deal with two classes of vectors, which are represented with a different notation: (i) positions ${\vec r}$ in a $d$-dimensional Euclidean space, in which the oscillators are embedded; (ii) state variables ${\mathbf X} = {(x_1, x_2, \ldots x_M)}^T$ in a $M$-dimensional phase space of the dynamical variables characterizing the state of the system at a given time $t$. There are $N$ oscillator cells located at discrete positions ${\vec r}_j$, where $j = 1, 2, \cdots N$, in the $d$-dimensional Euclidean space; and ${\mathbf X}_j$ is the state variable for each oscillator, whose time evolution is governed by the non-autonomous vector field ${\mathbf F}({\mathbf X}_j,t)$ [Fig. \ref{chemfig}]. The oscillators are not supposed to be identical, though, for they can have slightly different parameters. 

We suppose that the time evolution is affected by the local concentration of a chemical, denoted as $A({\vec r},t)$, through a time-dependent coupling function ${\mathbf g}$:
\begin{equation}
\label{oscil}
\frac{d{\mathbf X}_j}{dt} = {\mathbf F}({\mathbf X}_j,t) + {\mathbf g}(A({\vec r}_j,t)),
\end{equation}
\noindent meaning that the oscillator absorbs the chemical which secreted by all other oscillators and diffuses in the inter-oscillator medium. Hence the chemical concentration satisfies a diffusion equation of the form
\begin{equation}
\label{chem}
\frac{\partial A({\vec r},t)}{\partial t} + \eta A({\vec r},t) - D \nabla^2 A({\vec r},t) = \sum_{k=1}^{N} h({\mathbf X}_k) \delta({\vec r} - {\vec r}_k),
\end{equation}
\noindent where $\eta$ is a phenomenological damping parameter (representing the chemical degradation of the mediating substance) and $D$ is a diffusion coefficient. The diffusion equation above has a source term $h$ which depends on the oscillator state at the discrete positions ${\vec r}_j$: this means that each oscillator secrets the chemical with a rate depending on the current value of its own state variable. 

In Kuramoto's theory of coupling it is assumed that the diffusion is so fast, compared with the oscillator period, that we may set ${\dot A} = 0$ such that the concentration relaxes to a stationary value (adiabatic elimination). This is justifiable if the coupling time $\tau_c$ is much less than the dominant oscillator period $T$. Let $\ell$ be the characteristic length of the system: from the diffusion equation we can estimate the coupling time as $\tau_c \sim \ell^2/D$, where $D$ is the diffusion coefficient. Hence Kuramoto's model is valid provided $\tau_c \ll T$ or $ T \gg \ell^2/D$.

Given that the diffusion coefficient of most ions in water is $D \sim 10^{-9} m^2/s$ there follows that this approximation holds for $T \gg 10^9 \ell^2$, where $T$ is measured in seconds and $\ell$ in meters. Hence, unless $\ell$ is very short, this approximation is good for low-frequency phenomena. For example, the characteristic period of a spiking neuron is $T \sim 10 ms$, such that $\ell \ll 3 \times 10^3 m$, which is fairly true for any neuronal tissue. Circadian rhythms, for which $T \sim 24 h$ are also well described.

On the other hand, Kuramoto's model will be no longer valid if the timescales $T$ and $\tau_c$ are comparable, which can occur either if the coupling is slow or the oscillator dynamics is fast (high-frequency phenomena). In this case we have to solve the inhomogeneous diffusion equation (\ref{chem}) using Green's function method. In the general case there will be a boundary surface ${\cal S}$ on which we have to specify suitable boundary conditions. One such example is absorbing boundary conditions (Dirichlet), for which the concentration vanishes on ${\cal S}$, i.e. $A({\bf r} \in {\cal S},t) = 0$. Moreover we must specify the initial conditions, e.g. with a uniform initial concentration over the domain ${\cal D}$ bounded by ${\cal S}$: $A({\bf r},t=0) = {\cal A}({\bf r})$.

In this case the Green's function $G({\bf r},t;{\bf r'},t')$ satisfies the equation
\begin{equation}
\label{chemgreen}
\frac{\partial G}{\partial t} + \eta G - D \nabla^2 G = \delta({\vec r} - {\vec r'}) \delta(t-t'),
\end{equation}
with the following boundary and initial conditions $G({\bf r} \in {\cal S},t;{\bf r'},t') = 0$, $G({\bf r},t=0;{\bf r'};t') =0$, respectively. We also impose the causality condition $G({\bf r},t;{\bf r'};t') = 0$, if $ t < t'$.

Standard manipulations lead to the symmetry property of the Green' functions, $G({\bf r},t;{\bf r'},t')=G({\bf r'},-t';{\bf r},-t)$, and to  the solution of the inhomogeneous diffusion equation (\ref{chem})  \cite{butkov,arfken}
\begin{align}
\nn
 A({\vec r},t) & = \sum_{k=1}^{N} \int_0^{t^+} dt' h({\mathbf X}_k(t')) G({\bf r},t;{\bf r}_k,t') + \int_{\cal D} dV' A({\bf r'},t'=0) G({\bf r},t;{\bf r'},t'=0) + \\
\label{inomo}
 & D \int_0^{t^+} dt' \oint_{\cal S} dS' {\hat{\bf n'}}\cdot \left\lbrack \nabla' A({\bf r'},t') G({\bf r},t;{\bf r'},t') - \nabla' G({\bf r},t;{\bf r'},t') A({\bf r'},t') \right\rbrack,
\end{align}
where $t^+=t+ 0$ and $dS \, {\hat{\bf n}}$ denotes the vectorial area element on the bounding surface ${\cal S}$. After we find this concentration for any spatial point we make ${\vec r}={\vec r}_j$ for the $j$th oscillator, whose time evolution is then given by (\ref{oscil}):
\begin{equation}
\label{oscila}
\frac{d{\mathbf X}_j}{dt} = {\mathbf F}({\mathbf X}_j,t) + {\mathbf g}(A({\vec r}_j,t)),
\end{equation}

Among the most common coupling schemes we find: (i) the {\it linear coupling}, for which ${\mathbf g}(h({\mathbf X}_k)) = {\mathbf B}{\mathbf X}_k$, (ii) the {\it future coupling}, for which ${\mathbf g}(h({\mathbf X}_k)) = {\mathbf B}{\mathbf F}({\mathbf X}_k)$, (iii) the {\it nonlinear coupling}, for which ${\mathbf g}(h({\mathbf X}_k)) = {\mathbf B}{\mathbf H}({\mathbf X}_k)$, where ${\mathbf B}$ is a $M \times M$ matrix indicating which variables of the oscillators are coupled to whom, and ${\mathbf H}$ is a nonlinear function of its arguments. For example, using the linear coupling in (\ref{oscila}) gives us the following model
\begin{equation}
\label{osciladores}
\frac{d{\mathbf X}_j}{dt} = {\mathbf F}({\mathbf X}_j,t) + \sum_{k=1}^{N} \sigma({\vec r}_j - {\vec r}_k,t) {\mathbf B}{\mathbf X}_k.
\end{equation}

\section{Green function in the free space}

Let us begin by the simplest case in which there are no bounding surfaces, i.e. the domain is the entire space (this is equivalent to making the surface ${\cal S}$ ``going to infinity''), where the boundary condition is $A(|{\bf r}|\rightarrow\infty,t) = 0$. In this case the surface term in (\ref{inomo}) vanishes identically and the solution of the inhomogeneous equation is simply
\begin{equation}
\label{inomo1}
 A({\vec r},t) = \sum_{k=1}^{N} \int_0^{t^+} dt' h({\mathbf X}_k(t')) G({\bf r},t;{\bf r}_k,t') + \int dV' A({\bf r'},t'=0) G({\bf r},t;{\bf r'},t'=0).
\end{equation}

Yet for simplicity we can assume that the initial concentration is uniform throughout all space, such that $A({\bf r'},t'=0)=A_0$ and 
\begin{equation}
\label{inomo2}
 A({\vec r},t) = \sum_{k=1}^{N} \int_0^{t^+} dt' h({\mathbf X}_k(t')) G({\bf r},t;{\bf r}_k,t') + A_0 \int dV' G({\bf r},t;{\bf r'},t'=0).
\end{equation}

A further simplification can be made if the oscillator dynamics is considerably slower than the diffusion time (but not so slow that the concentration would relax immediately to its equilibrium value, as in Kuramoto's model), such that the term $h({\mathbf X}_k)$ is nearly constant over the interval $[0,t^+]$. In this case the concentration at any time approximately given by
\begin{equation}
 \label{conc}
 A({\vec r},t) \approx \sum_{k=1}^{N} h({\mathbf X}_k(t)) \sigma({\bf r}-{\bf r}_k,t) + A_0 N({\bf r},t),
\end{equation}
where we have defined the interaction kernel
\begin{equation}
 \label{kernel}
 \sigma({\bf r}-{\bf r}_k,t) = \int_0^{t^+} dt' G({\bf r},t;{\bf r}_k,t'),
\end{equation}
and the normalization integral (Note that, although $G({\bf r},t=0;{\bf r'},t')=0$, it turns out that $G({\bf r},t;{\bf r'},t'=0)$ is generally nonzero.)
\begin{equation}
 \label{norma}
 N({\bf r},t) = \int dV' G({\bf r},t;{\bf r'},t'=0).
\end{equation}

The Green's function of the $d$-dimensional free space is  
\beq
\label{invfour}
 G({\bf r},t;{\bf r'},t')  = \frac{1}{{(2\pi)}^{d}} \, H(t-t') e^{-\eta(t-t')} \int d^dk e^{i{\bf k}\cdot({\bf r}-{\bf r'}) - Dk^2 (t-t')}.
\eeq

For $d=1$ the Green function (\ref{invfour}) reads 
\begin{align}
 \label{green1}
 G(x,t;x',t') = \frac{H(t-t')}{\sqrt{4\pi D(t-t')}} \, e^{-\eta(t-t')} \exp\left\lbrack-\frac{{(x-x')}^2}{4D(t-t')}\right\rbrack,
\end{align}
In this case, the normalization integral (\ref{norma}) can be analytically solved to yield $N(x,t) = e^{-\eta t}$, showing that the influence of the initial condition $A_0$ is perceived only for relatively small times. Hence, if we are integrating over a time large enough it is justifiable to set $A_0 = 0$ for simplicity. 

The interaction kernel (\ref{kernel}) in this case is
\begin{equation}
 \label{kernel1}
 \sigma(x-x_k,t) = \int_0^{t^+} dt' \frac{e^{-\eta(t-t')}}{\sqrt{4\pi D(t-t')}} \exp\left\lbrack-\frac{{(x-x')}^2}{4D(t-t')}\right\rbrack,
\end{equation}
which, after a change of variables, reads
\begin{equation}
 \label{kernel01}
 \sigma(x-x_k,t) = \frac{x-x_k}{4D\sqrt{\pi}} \int_{u_1}^\infty du \frac{e^{-u-a/u}}{u^{3/2}},
\end{equation}
where 
\beq
 \label{adef}
 a = \frac{\eta{(x-x')}^2}{4D} = {\left(\frac{\gamma(x-x_k)}{2}\right)}^2, \qquad u_1 = \frac{{(x-x_k)}^2}{4Dt} 
\eeq
and we have defined a coupling length
\begin{equation}
 \label{length}
 \gamma = \sqrt{\frac{\eta}{D}}.
\end{equation}
The integral in (\ref{kernel01}) has to be numerically solved for each time $t$.

It is instructive to analyze the stationary limit of the interaction kernel (\ref{kernel01}), taking its $t\rightarrow\infty$ limit, for which $u_1\rightarrow 0$. There results
\begin{equation}
 \label{kernel01s}
 \sigma(x-x_k) = \lim_{t\rightarrow\infty} \sigma(x-x_k,t) = \frac{\gamma}{2\eta} \, e^{-\gamma(x-x_k)},
\end{equation}
which coincides, up to the normalization factor, with the earlier results of Kuramoto and coworkers \cite{kuramoto95,nakao99}, in their analysis of the fast-relaxation case.

For the two-dimensional case ($d=2$) the Green function (\ref{invfour}) becomes
\beq
\label{greend21}
G({\vec r},t;{\vec r'},t') = \frac{H(t-t')e^{-\eta(t-t')}}{4\pi D(t-t')}   \exp\left\lbrack - \frac{{(x-x')}^2+{(y-y')}^2}{4D(t-t')} \right\rbrack,
\eeq
and the normalization integral (\ref{norma}) is also $e^{-\eta t}$, as in the previous case. The interaction kernel is
\begin{equation}
 \label{kernel02}
 \sigma({\vec r}-{\vec r}_k,t) = \frac{1}{4\pi D} \int_{u_1}^\infty \frac{du}{u} e^{-u-a/u},
\end{equation}
where  
\beq
 \label{adef1}
 a = \frac{\eta[{(x-x_k)}^2+{(y-y_k)}^2]}{4D} = {\left\vert\frac{\gamma({\vec r}-{\vec r}_k)}{2}\right\vert}^2, \qquad 
 u_1 = \frac{{\vert{\vec r}-{\vec r}_k \vert}^2}{4Dt},
\eeq
which, in the stationary limit, reduces to the result already found by Nakao \cite{nakao99}:
\begin{equation}
 \label{kernel02s}
 \sigma({\vec r}-{\vec r}_k) = \lim_{t\rightarrow\infty} \sigma({\vec r}-{\vec r}_k,t) = \frac{1}{2\pi D} \, K_0[\gamma\vert{\vec r}-{\vec r}_k \vert].
\end{equation}

For the three-dimensional case ($d=3$) the Green function (\ref{invfour}) is
\beq
\label{greend31}
G({\vec r},t;{\vec r'},t') = \frac{H(t-t')e^{-\eta(t-t')}}{{[4\pi D(t-t')]}^{3/2}}   \exp\left\lbrack - \frac{{(x-x')}^2+{(y-y')}^2+{(z-z')}^2}{4D(t-t')} \right\rbrack,
\eeq
in such a way that the interaction kernel is
\beq
\label{sigma3d}
\sigma({\vec r}-{\vec r}_k,t) = - \frac{1}{4D \pi^{3/2}} \frac{1}{{|{\vec r}-{\vec r}_k|}} \int_{u_1}^\infty \frac{du}{\sqrt{u}} \exp\left(-u-\frac{a}{u}\right),
\eeq
where
\beq
 \label{adef3}
 a = \frac{\eta}{4D} {\vert{\vec r}-{\vec r}_k\vert}^2, \qquad u_1 = \frac{{\vert{\vec r}-{\vec r}_k \vert}^2}{4Dt},
\eeq
which, in the stationary limit, becomes
\beq
\label{sigma3ds}
\sigma({\vec r}-{\vec r}_k) = \lim_{t\rightarrow\infty} \sigma({\vec r}-{\vec r}_k,t) = \frac{1}{4D\pi} \frac{e^{-\gamma\vert{\vec r}-{\vec r}_k\vert}}{\vert{\vec r}-{\vec r}_k\vert}.
\eeq

\section{Green function in a finite domain}

If, instead of the free space, we consider that diffusion occurs only over a finite spatial domain, then we have a boundary value problem for the diffusion equation. A great simplification arises, however, if we consider absorbing boundary conditions, i.e. the concentration vanishes at the boundary surface ${\cal S}$. This kills the surface term in (\ref{inomo}) such that the solution of the inhomogeneous diffusion equation is formally similar to the free space case. 
\begin{equation}
\label{inh}
 A({\vec r},t) = \sum_{k=1}^{N} \int_0^{t^+} dt' h({\mathbf X}_k(t')) G({\bf r},t;{\bf r}_k,t'),
\end{equation}
where we set the initial condition as zero as well. This similarity stops here though, since the Green function depends in general on the boundary condition and hence the problem become mathematically more involved. One technique which can be used is the expansion in series of orthogonal eigenfunctions of a self-adjoint operator. We can express the delta functions in terms of these eigenfunctions and thus expand the Green function in series of them (bilinear formulas). In this section we will consider only the $\eta = 0$ case. 

A boundary value problem in one dimension consists of solving (\ref{chem}):
\begin{equation}
\label{chem1d}
\frac{\partial A}{\partial t} - D \frac{\partial^2 A}{\partial x^2} = \sum_{k=1}^{N} h({\mathbf X}_k) \delta(x - x_k),
\end{equation}
in a finite domain $0 \le x \le L$ with absorbing boundary conditions (Dirichlet) $A(x=0,t) = A(x=L,t) = 0$ and initial condition $A(x,t=0) = 0.$

The Green function of (\ref{chem1d}) is known by solving 
\begin{equation}
\label{chem1d}
\frac{\partial G}{\partial t} - D \frac{\partial^2 G}{\partial x^2} = \delta(x - x') \delta(t-t'),
\end{equation}
with $G(x=0,t;x',t') = G(x=L.t;x',t') = 0$, $G(x,t=0;x',t') = 0$.

The solution of (\ref{chem1d}) can be expressed as a bilinear formula involving the eigenfunctions $\phi_n(x)=\sqrt{2/L} \sin(k_n x)$, with corresponding eigenvalues $k_n=n\pi/L$: 
\beq
\label{greeneigen}
G(x,t;x',t') = \frac{2}{L} H(t-t') \sum_{n=1}^\infty \sin\left(\frac{n\pi x'}{L}\right) \sin\left(\frac{n\pi x}{L}\right) e^{-D n^2 \pi^2 (t-t')/L^2}.
\eeq
The interaction kernel can be expressed in terms of the eigenfunctions as 
\beq
\label{greeneigen}
\sigma(x_j-x_k,t) = \frac{2}{L} \sum_{n=1}^\infty \sin\left(\frac{n\pi x_j}{L}\right) \sin\left(\frac{n\pi x_k}{L}\right)\left( \frac{ 1 - e^{-D n^2 \pi^2 t/L}}{D n^2 \pi^2/L} \right).
\eeq

A two-dimensional boundary-value problem of interest is  
\begin{equation}
\label{chem2d1}
\frac{\partial A}{\partial t} - D \left( \frac{\partial^2 A}{\partial x^2} + \frac{\partial^2 A}{\partial y^2} \right) = \sum_{k=1}^{N} h({\mathbf X}_k) \delta(x - x_k) 
\delta(y - y_k),
\end{equation}
in a rectangular domain of sides $a$ and $b$ with absorbing boundary conditions (Dirichlet) 
\beq
\label{cc2}
A(x=0,y,t) = A(x=a,y,t) = A(x,y=0,t) = A(x,y=b,t) = 0,
\eeq
and initial condition $A(x,y,t=0) = 0$.

Taking into account the normalized eigenfunctions and the corresponding eigenvalues, there results that the Green function is given by the bilinear formula
\begin{align}
\nn
G(x,y,t;x',y',t') & = \frac{4 H(t-t')}{ab} \sum_{n=1}^\infty \sum_{m=1}^\infty \sin\left(\frac{n\pi x'}{a}\right)\sin\left(\frac{m\pi y'}{b}\right) \times \\
\label{G21}
& \sin\left(\frac{n\pi x}{a}\right)\sin\left(\frac{m\pi y}{b}\right)  \exp \left\{ -D \pi^2 \left(\frac{n^2}{a^2} + \frac{m^2}{b^2}\right)(t-t') \right\}
\end{align}
corresponding to the interaction kernel
\begin{align}
\nn
\sigma({\vec r}_j-{\vec r}_k,t) & = \frac{4}{ab D \pi^2} \sum_{n=1}^\infty \sum_{m=1}^\infty \sin\left(\frac{n\pi x_j}{a}\right)\sin\left(\frac{m\pi y_j}{b}\right) \times \\
\label{s21}
& \sin\left(\frac{n\pi x_k}{a}\right)\sin\left(\frac{m\pi y_k}{b}\right) {\left(\frac{n^2}{a^2} + \frac{m^2}{b^2}\right)}^{-1} (t-t') \left\lbrack
1 - \exp \left\{ -D \pi^2 t \left(\frac{n^2}{a^2} + \frac{m^2}{b^2}\right) \right\} \right\rbrack.
\end{align}

Another two-dimensional boundary problem consists of a circular domain of radius $r = a$ with an absorbing boundary. The diffusion equation in polar coordinates is 
\begin{equation}
\label{chem2dp}
\frac{\partial A}{\partial t} - D \left\lbrack \frac{1}{r} \frac{\partial}{\partial r} \left( r \frac{\partial A}{\partial r} \right) + \frac{1}{r^2} \frac{\partial^2 A}{\partial \theta^2} \right\rbrack = \frac{1}{2\pi r} \sum_{k=1}^{N} h({\mathbf X}_k) \delta(r - r_k) \delta(\theta - \theta_k) 
\end{equation}
where $0 \le r \le a$, $0 \le \theta < 2\pi$, with a Dirichlet boundary condition $A(r=a,\theta,t) = 0$ and initial condition $A(r,\theta,t=0) = 0$. Moreover the solution must be regular at the origin: $A(r=0,\theta,t) < \infty$.

From the eigenfunction expansion method, the Green function for this problem is \cite{duffy}
\beq
\label{greencirc1}
G(r,\theta,t;r',\theta',t') = \frac{1}{\pi D} \sum_{m=-\infty}^\infty \sum_{n=1}^\infty \frac{J_m\left(x_{mn} \frac{r}{a} \right) J_m\left(x_{mn} \frac{r'}{a} \right)}{{[{J'}_m(x_{mn})]}^2} \cos[m(\theta-\theta')] e^{-D x_{mn}^2 (t-t')/a^2},
\eeq
where $x_{mn}$ is the $n$th positive root of $J_m$. On integrating over time we obtain the corresponding interaction kernel
\beq
\label{greencirc1}
\sigma({\vec r}_j-{\vec r}_k,t) = \frac{1}{\pi a D} \sum_{m=-\infty}^\infty \sum_{n=1}^\infty \frac{J_m\left(x_{mn} \frac{r_j}{a} \right) J_m\left(x_{mn} \frac{r_k}{a} \right)}{{[{J'}_m(x_{mn})]}^2} \cos[m(\theta_j-\theta_k)] \left( 1 - e^{-D x_{mn}^2 t/a^2} \right).
\eeq

A complementary circular domain is all the plane minus a circle of radius $r = a$, whose boundary is absorbing. This can represent physically a ``sink''. The solution of the diffusion equation does not need to be regular at the origin, hence the general solution involves both Bessel and Neumann functions. The Green function is 
\beq
\label{gcc}
G(r,\theta,t;r',\theta',t') = \frac{1}{2\pi} \sum_{n=-\infty}^\infty \sum_{n=1}^\infty \cos[n(\theta-\theta')] H(t-t') \int_0^\infty d\alpha \alpha e^{-D\alpha^2(t-t')}  \frac{U_n(\alpha r)U_n(\alpha r')}{J_n^2(\alpha a) + N_n^2 (\alpha a)},
\eeq
where
\beq
\label{un}
U_n(\alpha r) = J_n(\alpha r) N_n(\alpha a) - J_n(\alpha a) N_n(\alpha r).
\eeq

If the domain is an annulus $a < r < b$, where $a$ is the internal and $b$ the external radius, the corresponding Green function is \cite{duffy}
\begin{align}
\nn
G(r,\theta,t;r',\theta',t') & = \frac{\pi H(t-t')}{4} \sum_{m=-\infty}^\infty \sum_{n=1}^\infty \alpha_{mn}^2 \frac{{[J_m(\alpha_{mn} r)]}^2 U_m(\alpha_{mn} r) U_m(\alpha_{mn} r')}{{[{J'}_m(\alpha_{mn} a)]}^2 - {[{J'}_m(\alpha_{mn} b)]}^2} \times \\
\label{greenanel}
& \cos[m(\theta-\theta')] e^{-D \alpha_{mn}^2 (t-t')},
\end{align}
where $\alpha_{mn}$ is the $n$th positive root of $U_n$, i.e. $U_n(\alpha_{mn} b) = 0$.

\section{Conclusions}

In a variety of problems in physical chemistry and cell biology the individuals, or cells, undergo some kind of dissipative dynamics leading to a stable limit-cycle as the asymptotic state. The dynamics along this limit-cycle can describe some physico-chemical or biological rhythm, and it can be mathematically described, in a minimal model, by a phase oscillator. We consider the situation in which the oscillator cells are coupled by the diffusion of some substance through the inter-cellular medium. This diffusing chemical is secreted and absorbed according to the cell dynamics, in such a way that the coupling is effectively mediated by the substance. 

In this paper we propose a general reaction-diffusion model for this process where we do not require the diffusion to be infinitely fast, what allows the possibility of diffusion at scales comparable to the oscillator characteristic period. The model can be formulated either in the free space (in one, two, and three spatial dimensions) or within closed domains, for which absorbing boundary conditions are assumed. Such boundary conditions, besides being simpler to impose, are actually necessary to avoid saturation of the diffusing substance in the inter-cellular medium. More complicated boundary conditions can be used, though, in order to tackle specific situations.

We give explicit expressions for linear, future and nonlinear couplings, in terms of an interaction kernel which is the time integral of the Green function corresponding to the diffusion equation for the appropriate Dirichlet problem. In some cases it is possible to evaluate this integral analytically, like when the degradation coefficient is zero. Otherwise the integral has to be done numerically. In closed domains the Green function is usually written as a converging sum which is also easily performed numerically whenever necessary.  

\section*{Acknowledgments}

This work has been partially supported by CNPq and CAPES (Brazilian Government Agencies). We acknowledge useful discussions and valuable comments by Tiago Kroetz.


\begin{thebibliography}{99}
\bibitem{winfree} A. T. Winfree, {\it The Geometry of Biological Time} (Springer Verlag, New York, 1980).
\bibitem{kuramoto95} Y. Kuramoto, Prog. Theor. Phys. {\bf 94}, 321 (1995). 
\bibitem{kuramoto96} Y. Kuramoto and H. Nakao, Phys. Rev. Lett. {\bf 76}, 4352 (1996)
\bibitem{kuramoto97} Y. Kuramoto and H. Nakao, Physica D {\bf 103}, 294 (1997).
\bibitem{battogtokh99} D. Battogtokh, Prog. Theor. Phys. {\bf 102}, 947 (1999).
\bibitem{nakao99} H. Nakao, Chaos {\bf 9}, 902 (1999).
\bibitem{kawamura07} Y. Kawamura, N. Nakao, and Y. Kuramoto, Phys. Rev. E {\bf 75}, 036209 (2007); 
\bibitem{battogtokh02} D. Battogtokh, Phys. Lett. A {\bf 299}, 558 (2002).
\bibitem{sakaguchi06} H. Sakaguchi, Phys. Rev. E {\bf 73}, 031907 (2006).
\bibitem{viana12} R. L. Viana, A. M. Batista, C. A. S. Batista, J. C. A. de Pontes, F. A. dos S. Silva, and S. R. Lopes, Commun. Nonlinear Sci. Numer. Simulat. {\bf 17}, 2924 (2012)
\bibitem{fabio16} F. A. dos S. Silva, S. R. Lopes, and R. L. Viana, Commun. Nonlinear Sci. Numer. Simulat. {\bf 35}, 37 (2016)
\bibitem{liu97} C. Liu, D. R. Weaver, S. H. Strogatz and S. M. Reppert, Cell {\bf 91}, 855 (1997).
\bibitem{liu00} C. Liu and S. M. Reppert, Neuron {\bf 25}, 123 (2000)
\bibitem{batista17} C. A. S. Batista, J. D. SzezechJr., A. M. Batista, E. E. N. Macau, and R. L. Viana, Physica A {\bf 470}, 236 (2017)
\bibitem{viana17} R. L. Viana, A. M. Batista, C. A. S. Batista, and K. C. Iarosz, Nonlinear Dyn. {\bf 87}, 1589 (2017).
\bibitem{veltman08} D. M. Veltman, I. Keizer-Gunnik, and P. J. M. Van Haastert, J. Cell Biol. {\bf 180}, 747 (2008).
\bibitem{cai12} H. Cai, C.-H. Huang, P. N. Devreotex and M. Iijima, Methods Mol. Biol. {\bf 757}, 451 (2012).
\bibitem{pikovsky} M. Rosenblum and A. Pikovsky, Phys. Rev. E {\bf 70}, 041904 (2004).
\bibitem{kuramoto} Y. Kuramoto, {\it Chemical Oscillations, Waves, and Turbulence} (Springer-Verlag, New York, 1984).
\bibitem{butkov} E. Butkov, {\it Mathematical Physics} (Addison Wesley, Reading, 1968).
\bibitem{arfken} G. B. Arfken e H. J. Weber, {\it Mathematical Methods for Physicists}, 5a. Ed. (Harcourt, San Diego, 2001).
\bibitem{duffy} D. G. Duffy, {\it Green's Functions with Applications} (Chapman \& Hall/CRC, Boca Raton, 2001).
\end{thebibliography}
\end{document}